\documentclass[apl,aip,  jmp,
amsmath, amssymb,   preprint,%
 reprint,%
]{revtex4-1}

\usepackage{graphicx}
\usepackage{dcolumn}
\usepackage{bm}

\begin{document}

\preprint{AIP/123-QED}

\title{$\pi$ junction transition in InAs self-assembled quantum dot coupled with SQUID }

\author{S. Kim}
\email[]{KIM.Sunmi@nims.go.jp} \affiliation{International Center for
Materials Nanoarchitectonics(MANA), National Institute for Materials
Science (NIMS), 1-1 Namiki, Tsukuba, Ibaraki 305-0044, Japan}

\author{R. Ishiguro }
\affiliation{Department of Applied Physics, Tokyo University of
Science, 1-3 Kagurazaka, Shinjuku Tokyo, 162-8601, Japan}

\author{M. Kamio}
\affiliation{Department of Applied Physics, Tokyo University of
Science, 1-3 Kagurazaka, Shinjuku Tokyo, 162-8601, Japan}

\author{Y. Doda}
\affiliation{International Center for Materials
Nanoarchitectonics(MANA), National Institute for Materials Science
(NIMS), 1-1 Namiki, Tsukuba, Ibaraki 305-0044, Japan}

\author{E. Watanabe}\author{D. Tsuya}
\affiliation{Nanotechnology Innovation Center, NIMS, 1-2-1 sengen,
Tsukuba, Ibaraki 305-0047, Japan}

\author{K. Shibata}
\affiliation{Institute of Industrial Science, University of Tokyo,
4-6-1 Komaba, Meguro-ku, Tokyo 153-8505, Japan}

\author{K. Hirakawa}
\affiliation{Institute of Industrial Science, University of Tokyo,
4-6-1 Komaba, Meguro-ku, Tokyo 153-8505, Japan}
\affiliation{CREST-JST, 4-1-8 Honcho, Kawaguchi, Saitama 332-0012,
Japan}

\author{H. Takayanagi}
\affiliation{International Center for Materials
Nanoarchitectonics(MANA), National Institute for Materials Science
(NIMS), 1-1 Namiki, Tsukuba, Ibaraki 305-0044, Japan}
\affiliation{Department of Applied Physics, Tokyo University of
Science, 1-3 Kagurazaka, Shinjuku Tokyo, 162-8601, Japan}
\affiliation{CREST-JST, 4-1-8 Honcho, Kawaguchi, Saitama 332-0012,
Japan}

\date{\today}

\begin{abstract}
We report the transport measurements on the InAs self-assembled
quantum dots (SAQDs) which have a unique structural
zero-dimensionality, coupled to a superconducting quantum
interference device (SQUID). Owing to the SQUID geometry, we
directly observe a $\pi$ phase shift in the current phase relation
and the negative supercurrent indicating $\pi$ junction behavior by
not only tuning the energy level of SAQD by back-gate but also
controlling the coupling between SAQD and electrodes by side-gate.
Our results inspire new future quantum information devices which can
link optical, spin, and superconducting state.

\end{abstract}

\maketitle

%


Recent advances in nanofabrication technology have made it possible
to couple quantum dot (QD) with superconducting quantum interference
device (SQUID), so that the combination of highly controllable
electronic system and most sensitive detector for magnetic flux
inspires new applications for future quantum information device. Up
to date, the QD-SQUID has been studied by utilizing one-dimensional
semiconductor nanostructures such as carbon
nanotubes\cite{Cleuziou:2006} and InAs nanowires \cite{Dam:2006}
where the gate-controlled $\pi$ junction suggests that such
configuration is very attractive for studying the spin states of an
individual electron placed on the one of the two QD Josephson
junctions in SQUID.

Besides one-dimensional nanostructure, it is noticeable for the
recent researches of the self-assembled quantum dot (SAQD) as
another type of QD. The SAQDs are grown by strain driven
self-assembly onto a substrate, leading to spontaneous formation of
small islands. Since it can be optically excited, controllably
positioned, electronically coupled, and embedded into active
devices, it is very attractive for the realization of optically
programmable electron spin memory.\cite{Kroutvar:2004} Also note
that the geometric structure can have better flexibility than
one-dimensional nanostructure for complicated-circuit integration.
Recently InAs SAQD has been coupled with superconductor via the
nanogap method \cite{Jung:2005} and has been studied for the
interplay between Kondo effect and superconductivity in
SAQD.\cite{Buizert:2007, Deacon:2010} In spite of its versatility,
it has not been yet reported for SAQD coupled to SQUID.

In this letter, we present the successful fabrication for the InAs
SAQD coupled with SQUID by means of an electron beam (EB)
lithography and an atomic force microscope (AFM) and the electrical
transports involving gate-controlled $\pi$ junction transition of
QD-SQUID.

\begin{figure}
\includegraphics[width=8.5cm]{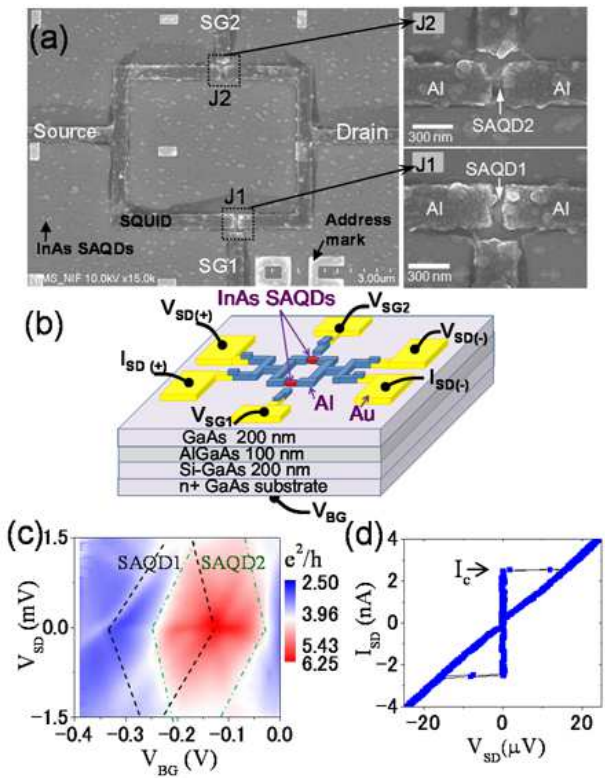}
\caption{\label{fig:aaa} (a) SEM image of the SQUID coupled with
InAs SAQD. Here randomly distributed small dots are InAs SAQDs. For
two junctions of Al/SAQD/Al marked as dotted square (J1 and J2), the
enlarged images are shown in the right insets. (b) Schematic diagram
of our device. (c) Differential conductance $dI/dV$ as functions of
source-drain voltage ($V_{SD}$) and back-gate voltage ($V_{BG}$)
under the magnetic field of 100 mT and at $V_{SG1}$=0 V. (d) The
$I-V$ characteristics of SQUID at about 30 mK and in an absence of
magnetic field. Here $V_{SG1}$ was set to be -0.04 V and $V_{BG}$
grounded. }
\end{figure}

In our experiment, the InAs SAQDs with diameters 100$\sim$250 nm
were grown by molecular beam epitaxy with a sequential deposition
technique on  n$^+$-GaAs substrate.\cite{Shibata:2007} A schematic
diagram of our device is shown in Fig.~\ref{fig:aaa}(b). The
electrodes were prepared by two parts, i.e., outer Au electrodes (5
nm Ti /250 nm Au) by means of maskless laser lithography and inner
Al electrodes (5 nm Ti /100 nm Al) by EB lithography with a high
positioning accuracy. The Al electrodes are directly contacted to
uncapped SAQD as superconducting electrodes. The positioning
accuracy was less than 10 nm in best condition which is enough to
make contact on such a small island of SAQD. All metal layers were
prepared by electron beam deposition. Before fabricating Al
electrodes, the address marks (5 nm Ti/30 nm Pt) were patterned by
the EB lithography and then the AFM mapping was carried out to know
the SAQDs location. Also, in order to deoxidize the dot surface and
have highly transparent interfaces for an observable supercurrent,
the SAQDs surface was etched by buffered hydrofluoric acid (50:1)
for 30 s. The normal resistance of SQUID was measured to be about 3
k$\Omega$ at room temperature.

All transport measurements were performed in a dilution refrigerator
with a base temperature $\sim$30 mK which is well below the
superconducting transition temperature of our sample $T_c\sim$1 K.
Several filtering systems consisting of the mini-circuit $LC$
filters at room temperature, the Thermocoax,\cite{Zorin:1995} and
the shielded $RC$ filters at mixing chamber covering all frequencies
above about 1 KHz were employed to minimize the external noise.

Figure~\ref{fig:aaa}(a) shows a scanning electron microscope (SEM)
image of the fabricated QD-SQUID employing SAQD. The Al-based
superconducting loop has 300 nm width and loop size of
3.87$\times$3.08 $\mu$m$^2$. The gap where SAQD was embedded is
about 100 nm and  two Al/SAQD/Al junctions in SQUID are denoted as
J1 and J2. From the AFM image which is taken before deposition of Al
electrodes, it is known that the SAQDs have the elliptical shapes
with short/long diameters i.e., 140 nm/200 nm for one in J1 and 160
nm/220 nm in J2. Indicated as SG1 and SG2, two side-gates are
located 100 nm far from the SQUID loop for individual control of the
electron state in SAQD, which is contrast with a back-gate (BG)
which tunes two SAQDs simultaneously. Unfortunately the available
gates for the presented sample were only for SG1 and BG, because
there was the electrical connection between SG2 and BG by a possible
micro-defect.

In order to characterize the QD properties, we measured a
differential conductance (versus a source-drain voltage $V_{SD}$ and
a back-gate voltage $V_{BG}$) under the magnetic field of 100 mT to
suppress superconductivity and at $V_{SG1}$=0 V (see
Fig.~\ref{fig:aaa}(c)). Note that since the $V_{BG}$ simultaneously
changes the properties of two QDs (SAQD1 and SAQD2) which are
parallelly connected in SQUID, Fig.~\ref{fig:aaa}(c) depicts the
mixture of the Coulomb diamonds of two QDs. We mark simple guide
lines for the Coulomb diamonds corresponding to odd occupation of
each dot in the Fig.~\ref{fig:aaa}(c) where dot property for SAQD1
is distinguishable due to $V_{SG1}$ response (not shown). We roughly
estimate the dot properties in SQUID to be a charging energy
$U$$\sim$1.7 meV and a level spacing $\delta$$E$$\sim$0.9 meV for
SAQD1 while it is too vague to assign for SAQD2. The Coulomb
oscillations with $V_{BG}$ have been detected with the differential
conductance ranging up to 6.25 $e^2/h$. By changing $V_{SG1}$, the
conductance value of SAQD1 only is shifted with a value of less than
2 $e^2/h$. Also the strong Kondo effect was observed for SAQD2 due
to strong dot-lead coupling. We think that the high conductance over
6$e^2/h$ is mainly contributed from SAQD2 being in open dot regime.

\begin{figure}
\includegraphics[width=8.5cm]{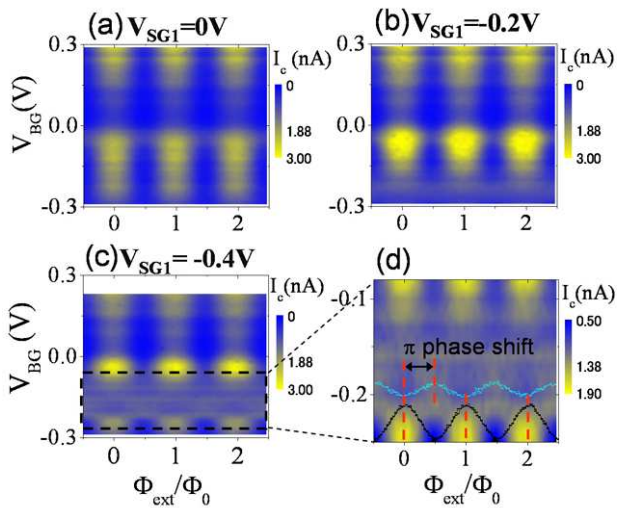}
\caption{\label{fig:bbb} $I_c$ oscillation as a function of external
magnetic field and $V_{BG}$. $V_{SG1}$ is fixed to be 0V for (a),
-0.2 V for (b), and -0.4 V for (c). (d) is enlarged plot of (c) in
the range of $V_{BG}$ from -0.08 V to -0.25 V. For clarity, line
profiles of $V_{BG}$=-0.2 V (blue) and $V_{BG}$=-0.23 V (black) are
superimposed with an arbitrary scale. }
\end{figure}

In an absence of any magnetic field, the $I-V$ characteristic of
SQUID exhibits a clear supercurrent (with a critical value of
$I_c$=2.5 nA) with accompanying a hysteretic behavior as shown in
fig.~\ref{fig:aaa}(d). As a function of external magnetic field
($\Phi_{ext}$), the periodic $I{_c}$ oscillation is shown in
Fig.~\ref{fig:bbb}(a) where the period of about $\Phi_0$=1.5 gauss
is well consistent with the addition of a flux quantum $\phi_0=h/2e$
to the effective SQUID area. This result implies that the loop
including Al/SAQD/Al junctions operate as SQUID. For individual
control of SAQD1, we additionally tuned $V_{SG1}$ to be -0.2 V
(Fig.~\ref{fig:bbb}(b)) and -0.4 V (Fig.~\ref{fig:bbb}(c)).
Interestingly we observed $\pi$ junction behavior for $V_{SG1}$=-0.4
V which manifests as $\pi$ phase shift on $I{_{c}}$ oscillation at
certain voltage regime of $V_{BG}$ (see Fig.~\ref{fig:bbb}(d) which
is enlarged graph of (c)). This means the presence of spontaneous
supercurrent in the loop induced by control of $V_{SG1}$ and only
one of SAQD junctions is in $\pi$ junction state while the other is
so-called 0 junction.

\begin{figure}
\includegraphics[width=9cm]{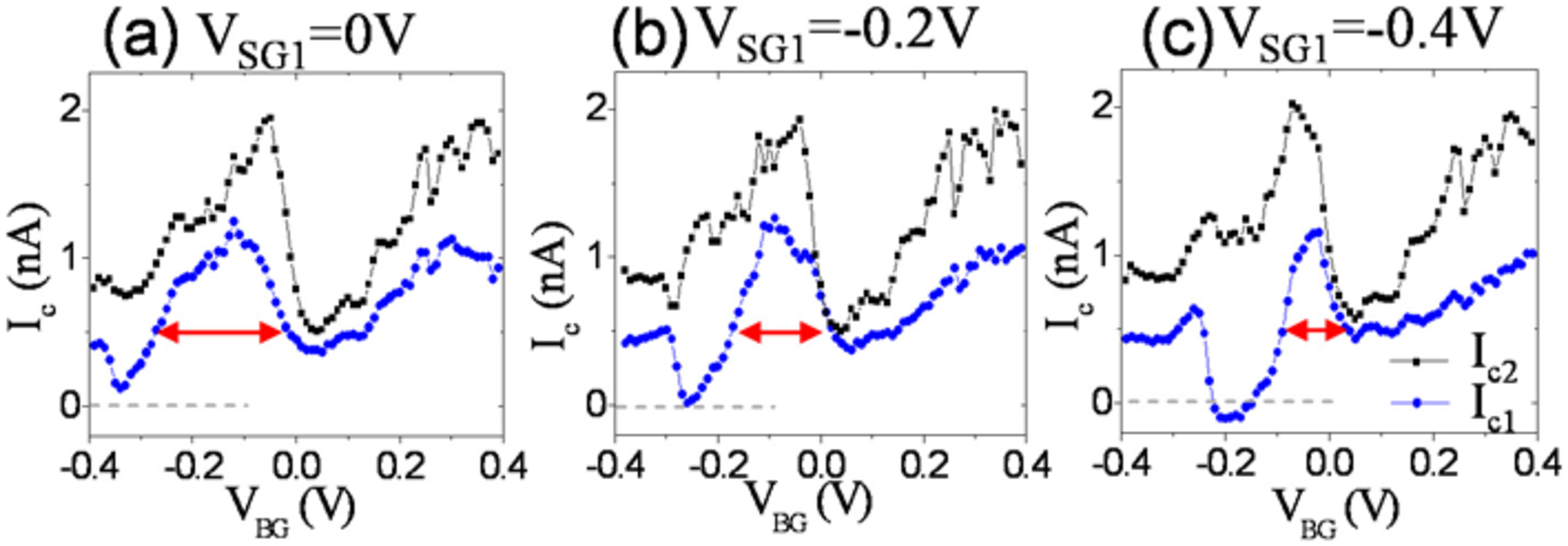}
\caption{\label{fig:ccc}  $I_c$ profiles of individual junctions in
SQUID with varying $V_{BG}$ at fixed $V_{SG1}$=(a) 0 V, (b) -0.2 V,
(c) -0.4 V.
 }
\end{figure}

 For more detailed study for $\pi$-junction transition, we plot the
individual $I{_c}$ profiles of each junction in Fig.~\ref{fig:ccc}
which is resultant from the analysis of the interference properties
of SQUID in the limit of small self inductance as described below.
This method is very useful when we want to know the individual
$I{_c}$ properties for the QD-SQUID where more than one of junctions
in the SQUID can not be (Coulomb) blockaded, for instance open dot
case.

Due to a relatively small $I_c$ and small self inductance of about
15 pH calculated from SQUID geometry, the total $I_{c}$ of our
device can be expressed as the following expression in the limit of
small self inductance \cite{Barone:p371}
\begin{eqnarray}
I_c(\Phi_{ext})=[I_{c1}^{2}+I_{c2}^{2}+2I_{c1}I_{c2}\cos2\pi(\Phi_{ext}/\Phi_0)]^{1/2}
 \label{eq:1}
\end{eqnarray}
where $I_{c1}$ and $I_{c2}$ are the critical currents of each
junction, J1 and J2. According to Eq.~(\ref{eq:1}), $I_c$ amplitude
has maximum corresponding to $|I_{c1}$+$I_{c2}|$ at
$\Phi_{ext}$=$\Phi_0$ and minimum being $|I_{c1}$-$I_{c2}|$ at
$\Phi_{ext}$=$\Phi_0/2$. Thus we can obtain $I_{c2}$ from
$\{$$I_c$($\Phi_0$) -$I_c$($\Phi_0/2$)$\}$/2 and also $I_{c1}$ from
$I_{c}$($\Phi_0$)-$I_{c2}$.

As shown in Fig.~\ref{fig:ccc}, we obtained only positive values for
the $I_{c2}$ which implies that the 0-$\pi$ junction transition does
not occur in J2. Only for J1, we found a negative supercurrent at
specific voltage range which corresponds to the region showing $\pi$
phase shift on $I_c$ oscillation in Fig.~\ref{fig:bbb}(d).
Such supercurrent reversal has been demonstrated by adding a single
electron to the QD Josephson junctions using carbon
nanotube\cite{Cleuziou:2006, Jorgensen:2007} and InAs
nanowire.\cite{Dam:2006} Also other mechanisms for negative
supercurrent have been experimentally studied in d-wave high $T_c$
superconductor,\cite{Harlingen:1995} a
superconductor/ferromagnet/superconductor,\cite{Ryazanov:2001} and a
superconductor/normal metal/superconductor Josephson
junction.\cite{Baselmans:1999}

In our experiment, the $V_{BG}$ is used to tune mainly the energy
level in the quantum dots and the $V_{SG1}$ is for the change of
coupling between QD1 and superconducting leads. In
Fig.~\ref{fig:ccc}(a)-(c), the $I_c$ maxima are induced by the
resonance between the energy levels in SAQD and the Fermi energy in
the superconductor and the broadening of the peaks depends on the
dot-lead coupling. One can notice that the peak widths of $I_{c1}$
indicated by the arrows in Fig.~\ref{fig:ccc} are changed by the
$V_{SG1}$. This fact implies that the SG1 controls the coupling
energy. This is good agreement with Ref. \onlinecite{Kanai:2010}
where although there is no direct observation of $\pi$-junction
behavior, it shows the possibility of the phase transition between
Kondo singlet and magnetic doublet ground states driven by the
side-gate modulation of the device tunnel coupling.

In summary, we fabricated a SAQD-SQUID which is the first try in a
sense of combination of the uncapped InAs SAQD with unique
structural zero-dimensionality and SQUID. Using SQUID geometry, we
directly observed the $\pi$ junction behavior accompanying negative
supercurrent by tuning a side-gate. Our results support
Ref.\onlinecite{Kanai:2010} where the $\pi$ junction transition was
explained by the singlet-doublet transition of the spin state of the
InAs SAQD due to a change of the coupling between SAQD and
superconducting leads by side-gate without direct observation of
$\pi$ state. Employing InAs SAQD to QD-SQUID may open new
possibility for a quantum information processing in a wide range of
optics, spintronics, and superconductivity.

This research is supported by the Japan Society for the Promotion of Science(JSPS)
 through its FIRST program and Grant-in-Aid for Scientific Research(KAKENHI) S (No. 20221007).
The author (S. Kim) would like to thank to R. Inoue and S. Moriyama
for helpful discussions.

\end{document}